\documentclass[twocolumn,showpacs,preprintnumbers,amsmath,amssymb]{revtex4}

\usepackage{graphicx}
\usepackage{dcolumn}
\usepackage{bm}
\begin{document}

\preprint{APS/123-QED}

\title{Astrophysical inputs on the SUSY dark matter annihilation detectability}

\author{F. Prada$^{1}$, A. Klypin$^2$, J. Flix$^3$, M. Mart\'\i nez$^3$ and
E. Simonneau$^4$} 
\medskip
\affiliation{$^1$ Ram\'on y Cajal Fellow, Instituto de Astrof\'\i sica de Andaluc\'\i a (CSIC), E-18008 Granada, Spain  \\ 
$^2$ Department of Astronomy, New Mexico State University, Las Cruces
NM 88003-8001, USA \\
$^3$ Institut de F\'\i sica D'Altes Energies, Universitat Autonoma, 
E-08193 Barcelona, Spain \\
$^4$ Institut d'Astrophysique de Paris, CNRS, 75014 Paris, France}


\date{January 8, 2004}

\begin{abstract}
If dark matter (DM), which is considered to constitute most of the
mass of galaxies, is made of supersymmetric (SUSY) particles, the centers
of galaxies should emit $\gamma$-rays produced by their self-annihilation.  
We present accurate estimates of continuum 
$\gamma$-ray fluxes due to neutralino annihilation in the
central regions of the Milky Way. We use detailed models of our Galaxy, 
which satisfy available observational data, and include some 
important physical processes, which were previously neglected. 
Our models predict that spatially extended annihilation signal should be 
detected at high confidence levels by incoming experiments 
assuming that neutralinos make up most of the DM in the 
Universe and that they annihilate according to current SUSY
models.

\end{abstract}

\pacs{98.80.-k, 98.35.Ce, 98.35Gi, 95.35.+d,14.80.Ly}
\maketitle

There is an increasing hope that the new generation of Imaging 
Atmospheric $\check{C}$erenkov Telescopes (IACTs) would detect in 
the very near future the $\gamma$-ray signal coming from the annihilation 
products of the SUSY DM in galaxy halos 
(e.g., \cite{Bergstrom98,Baltz00,TO,Stoehr}). The success 
of such a detection in competition with other indirect or direct 
experiments including accelerators will solve one of the most fundamental
questions in Astrophysics and Particle Physics: the nature of the 
dark matter. The lightest supersymmetric particle (LSP) has been proposed to
be a suitable candidate for the non-baryonic cold DM
(\cite{Goldberg,Ellis84}, see also \cite{Primack,Jungman} for 
a review). The LSP is stable in SUSY models
where R-parity is conserved \cite{Weinberg,HS,Allanach} and its 
annihilation cross section and mass has the appropriate 
relic densities \cite{Jungman,Edsjo03} in the range 
allowed by WMAP, i.e. $0.095 < \Omega_{CDM} h^2 < 0.129$ 
\cite{Spergel}. We focus in the Minimal 
Supersymmetric extension of the Standard Model of particle physics (MSSM) 
where the LSP is the neutralino ($\chi$). New upper limits on the 
neutralino mass ($m_\chi$)
have been estimated due to the constrains on the neutralino relic 
densities $\Omega_\chi h^2$ provided by WMAP; $m_\chi < 1500 \,$GeV
based on the  MSSM including  minimal supergravity
(mSUGRA) \cite{Ellis03,Edsjo03}. A lower limit of $m_\chi \sim 100\,$GeV has
been set by the accelerators \cite{Hagiwara}. 

The number of neutralino annihilations in galaxy halos and therefore
the expected gamma signal arriving at the Earth depends not only on
the adopted SUSY model but also strongly depends on the DM
density $\rho_{dm}(r)$. This is why the central region $r< 200\,$pc of
the Milky Way, where  the density is the largest, is the favorite site to
search for this signal. The expected total number of continuum $\gamma$-ray 
photons received per unit time and per unit area, from a circular 
aperture on the sky of width $\sigma_{\rm t}$ (the resolution of the 
telescope) observing at a given direction $\Psi_0$ relative  to 
the center of the Milky Way can be written as:
\begin{eqnarray}
F(E>E_{\rm th})=\frac{1}{4\pi} {\rm YSUS} \cdot U(\Psi_0), 
\label{eq:one}
\end{eqnarray}
$${\rm YSUS}= \frac{N_{\gamma} \left<\sigma v\right>}{2 m_\chi^2}, \: U(\Psi_0)=\int J(\Psi)B(\Omega)d\Omega,$$

where the factor YSUS  (back-spelled SUSY)
 depends only on the physics of annihilating particles and all the
 astrophysical properties (such as the DM distribution and 
geometry considerations) 
appear only in the factor $U(\Psi_0)$.
This factor also  accounts for the beam smearing, where
$ J(\Psi) =  \int_{l.o.s} \rho_{dm}^2(r) \, dl $,
 $dl = \pm r dr/{\sqrt{r^2-d_{\odot}^2 \sin^2\Psi}},$
is the integral of
the line-of-sight of the square of the DM density 
along the direction $\Psi$,
and $B(\Omega)d\Omega$ is the Gaussian beam of the telescope:
\begin{equation}
B(\Omega) d\Omega  =  \exp\left[ -\frac{\theta^2}{2\sigma_t^2}\right] 
sin\theta \, d\theta d\varphi .
\end{equation}
 The angles $\theta$ and $\varphi$ are related with the direction of
observation $\Psi_0$ and the line-of-sight angle $\Psi$ by $\cos\Psi =
\cos\Psi_0 \cos\theta + \sin\Psi_0 \sin\theta \cos\varphi$.  We have
assumed spherical symmetry for the DM particles around the
Galactic Center and that the observer is located in the Calactic
equatorial plane at a distance $d_{\odot}$ (here 8.0 kpc).

The factor YSUS/$4\pi$ represents the isotropic
probability of $\gamma$-ray production per unit of DM 
density. It can be determined for any SUSY
model  given the neutralino mass $m_\chi$, the number of continuum
$\gamma$-ray photons $N_{\gamma}$ emitted per annihilation, with
energy above the IACT energy threshold ($E_{th}$), and  the
thermally average cross section $\left<\sigma v\right>$ of the
DM particles. We can then estimate a YSUS parameter range, given
a neutralino mass interval of $100-1500\,$GeV, and a  
cross section $\left<\sigma v\right>$ interval of 
$5\times10^{-27}-3\times10^{-26}$ $cm^3 s^{-1}$ obtained for 
a sample of MSSM models computed in \cite{Stoehr, Tasi} with 
relic densities in agreement with the WMAP constrains. The number of  
continuum gamma photons produced per annihilation N$_{\gamma}$ 
is obtained by integrating the continuum spectrum given by the decay 
of $\pi^0$ mesons produced in the fragmentation of
quarks. It can be well approximated by the eq.(18) in 
\cite{TO}, i.e. $N_{\gamma} = 5/6 \, x^{3/2}-10/3 \, x+5 \,\sqrt{x}+5/(6 \,\sqrt{x})-10/3$, where $x\equiv E_{\rm th}/m_\chi$. This gives 
values of the YSUS parameter in the range of $10^{-34}$
to $10^{-30}$  photons GeV$^{-2} cm^3 s^{-1}$ for $E_{th}$ from 1 to 400\,GeV.

A cuspy DM halo 
$\rho_{dm}(r) \propto$ r$^{-\alpha}$
predicted by the simulations of the Cold Dark Matter with the
cosmological constant ($\Lambda$CDM ) is often assumed for the
calculations of $U(\Psi_0)$ (e.g.,
\cite{Bergstrom98,Baltz00,TO,Stoehr,Bergstrom99,Calcaneo,UBEL,HD,TS,Tasi,Evans}).
Cosmological $N-$body simulations indicate that the distribution of DM
in relaxed halos varies between two shapes: the NFW \cite{NFW} density
profile $\rho(r) =\rho_0/x(1+x)^2,$ $x\equiv r/r_s$ with asymptotic
slope $\alpha = 1$ and the steeper Moore et al. \cite{Moore98} profile
$\rho(r) =\rho_0/x^{1.5}(1+x)^{1.5}$, $\alpha = 1.5$. The density of
DM also depends on two parameters of the approximations: the virial
mass $M_{\rm vir}$ and the concentration $C \equiv r_{\rm vir}/r_s$,
where $r_s$ is the characteristic radius of assumed approximation. For
Milky Way-size halos the average concentration $C=15$ and the
$1\sigma$-variance is $\Delta\log(C)=0.11$. For Moore et al profile 
we define concentration as $C_{moore}=C_{\rm NFW}*1.72$.

{\it Milky Way mass models with adiabatic compression.---} The predictions for the DM halos are valid {\it only} for
halos without baryons.  When normal gas (``baryons'') loses its energy
through radiative processes, it falls to the central region of forming
galaxy.  As the result of this redistribution of mass, the
gravitational potential in the center changes substantially. The dark
matter must react to this deeper potential by moving closer to the
center and increasing its density. This increase in the DM density is
often treated using adiabatic invariants. This is justified because
there is a limit to the time-scale of changes in the mass
distribution: changes of the potential at a given radius cannot happen
faster than the dynamical time-scale defined by the mass inside the
radius.  Adiabatic contraction of dark matter in a collapsing protogalaxy was
used already in 1962 \cite{ELS}.  In 1980, 
Zeldovich et al. \cite{Zeldovich}
used it to set constraints of properties of elementary particles
(annihilating massive neutrinos). They were also the
first to present analytical expression for adiabatic compression
(for pure radial orbits) and to make numerical tests to confirm that
the mechanism works. The present form of analytical approximation
(circular orbits) was introduced in \cite{Blumenthal}. 
If $M_{\rm in}(r_{\rm in})$ is the initial distribution of mass (the one
predicted by cosmological simulations), then the final (after
compression and formation of galaxy) mass distribution is given by
$M_{\rm fin}(r)r = M_{\rm in}(r_{\rm in})r_{\rm in}$, where $M_{\rm
fin} =M_{\rm DM} + M_{\rm bar}$. This approximation was tested in
numerical simulations \cite{White85, JNB}.
The approximation assumes that
orbits are circular and, thus $M(r)$ is the mass inside the
orbit. This is not true for elongated orbits: mass $M(r)$ is 
smaller than the real mass, which a particle ``feels'' when it
travels along elongated trajectory. This difference in masses requires
a relatively small correction: mass $M$ should be replaced by the mass
inside time-averaged radius of trajectories passing through given
radius $r$: 
$M_{\rm fin}(\left<r\right>)r = M_{\rm in}(\left<r_{\rm in}\right>)r_{\rm
in}$. We find the correction using Monte Carlo realizations of
trajectories in the NFW equilibrium halo and finding the time-averaged
radii $\left<x\right>\approx 1.72x^{0.82}/(1+5x)^{0.085}$, $x\equiv r/r_s$. 
This approximation predicts smaller contraction in the central regions,
where individual trajectories are very elongated. It gives better fits
than the standard approximation when compared with realistic
cosmological simulations \cite{Kravtsov2004}.

\begin{table}
\caption{\label{tab:table1}Models and constraints for the Milky Way Galaxy}
\begin{ruledtabular}
\begin{tabular}{lccc}
   & Model A    & Model B  & Constr.    \\ 	
   & NFW & Moore et al.  \\ \hline
Virial mass, $10^{12}M_{\odot}$  & 1.07   & 1.14 & -- \\
Virial radius, kpc   &  264  & 270  & --\\
Halo concentration C  & 11     &  12   & 10.3-21.5    \\
  &      &     &  ($1.5\sigma$)   \\
Disk mass, $10^{10}M_\odot$ & 3.7 & 4.0 & -- \\
Disk scale length, kpc & 3.2 & 3.5 & 2.5-3.5 \\
Bulge mass, $10^9M_{\odot}$ & 8.0  & 8.0 & -- \\
Black Hole mass, $10^6M_\odot$  & 2.6 & 2.6 & 2.6 \\
$M(<100{\rm kpc})$, $10^{11}M_\odot$  & 6.25 & 5.8 & $7.5\pm 2.5$ \\
$\Sigma_{\rm total}$, $|z|<1.1$~kpc  &65  &70 & $71\pm 6$ \\
\qquad  at $R_{\odot}$, $M_{\odot}{\rm pc}^{-2}$ \\
$\Sigma_{\rm baryon}$ at $R_{\odot}$,$M_{\odot}{\rm pc}^{-2}$ &47  &53 & $48\pm 8$ \\
$V_{\rm circ}$ at 3~kpc, km/s &203 &205& $200\pm 5$\\
\end{tabular}
\end{ruledtabular}
\end{table}

\begin{figure}
\includegraphics[width=.45\textwidth]{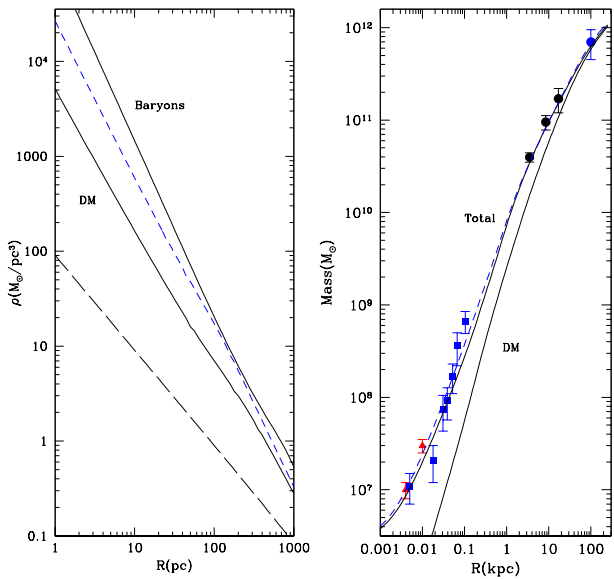}
\caption{\label{fig:fig1} Density and mass profiles for the Milky Way
Galaxy.  Symbols on right panel show observational constraints as
taken from Klypin et al. \cite{Klypin02}.  The  full and dashed curves labeled "Total" are total mass in NFW and Moore et al. models with adiabatic
compression.  DM mass in the NFW model is shown by the
thick curve. In the central region most of the mass is in
baryons. Left panel shows the density. The top full curve is the density of baryons. The dashed and full curves labeled "DM" are for the Moore
et al. and NFW  models with adiabatic compression. The long-dashed curve is the
uncompressed NFW profile for comparison.}
\end{figure}

In order to make realistic predictions for annihilation rates, we
construct two detailed models of the Milky Way Galaxy by redoing the
full analysis of numerous observational data collected in
\cite{Klypin02}. The models are compatible with the available
observational data for the Milky Way and their main parameters are
given in Table~\ref{tab:table1}. More details on the model ingredients
and the existing observational constrains can be found in
\cite{Klypin02}.  Fig.~\ref{fig:fig1} presents the distribution of
mass and density in the models. While all observations were included,
some of them are more important than others.  The solar neighborhood is
relatively well studied and, thus, provides important observational
constraints. In Table~\ref{tab:table1} we present two local
parameters: the total density of matter inside 1.1~kpc $\Sigma_{\rm
total}$ (obtained from kinematics of stars) and the surface density of
gas and stellar components $\Sigma_{\rm baryon}$. Circular velocity
$V_{\rm circ}$ at 3~kpc distance from the center provides another
crucial constraint on models as emphasized in \cite{BinneyEvans}. It
is difficult to estimate errors of this parameter because of uncertain
contribution of the galactic bar. We use $\pm 5$km/s error, which is
realistic, but it can be even twice larger. Probably the most debated
constraint is coming from counts of microlensing events in the
direction of the galactic bulge. Our models are expected to have 
the optical depth of microlensing events $\tau =1.2-1.6\times10^{-6}$ 
and, thus, they are compatible with the 
values of $\tau$ determined recently from the observations $\tau =1-1.5\times
10^{-6}$ \cite{AfonPopo}, but are excluded if $\tau > 2\times 10^{-6}$ 
(see \cite{Klypin02} for a detailed discussion on the bulge
optical depth in our models).


\begin{figure}
\includegraphics[width=.45\textwidth]{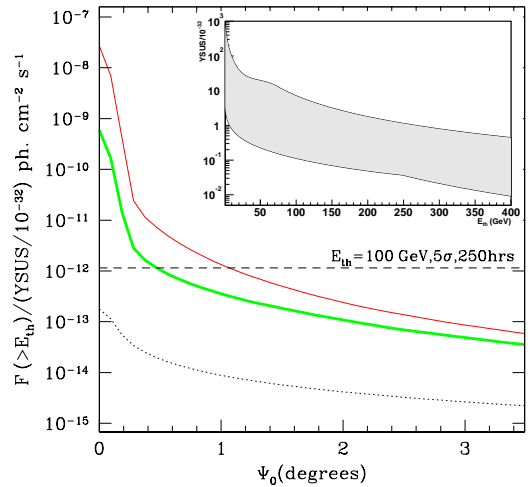}
\caption{\label{fig:fig2} Predicted continuum gamma flux 
as a function of distance $\Psi_0$ from the Galactic Center for
Models A and B. The thick line shows the flux for the
compressed NFW DM density profile of the Model A and the thin line for
the compressed Moore et al. profile of the Model B. The flux 
profile for the uncompressed NFW profile is also shown for
comparison (dotted line). The dashed line give the minimum detectable 
gamma flux F$_{\rm min}$ at $5\sigma$ level for exposure of 250 hours 
and E$_{\rm th}$=100\,GeV for a typical IACT. The inserted pannel shows
the YSUS/$10^{-32}$ dependence with the IACT E$_{\rm th}$. For a given
E$_{\rm th}$, the shadow region scans all the $m_\chi$, 
$\left<\sigma v\right>$ and N$_\gamma$ intervals (see text).}
\end{figure}

{\it Gamma-ray annihilation observability in the Milky Way.---} The expected neutralino annihilation gamma flux, in units 
of YSUS/$10^{-32}$, can be computed from Eq. 1 for the compressed DM 
density profile provided by our
Milky Way models as a function of the angular distance $\Psi_0$ from
the Galactic Center. In Fig.~\ref{fig:fig2} we show the predicted
fluxes. We also show as a comparison the expected flux for the
uncompressed NFW density profile. The flux profiles were determined
for a typical IACT of resolution $\sigma_t=0.1^\circ$ and solid angle
$\Delta\Omega=10^{-5}sr$. We have multiplied
the flux profiles by a factor of 1.7 quoted by Stoehr et al. \cite{Stoehr} 
to account for the presence
of substructure inside the Milky Way halo  \cite{Klypin99,Moore99}.

The success of a detection requires that the minimum detectable gamma
flux F$_{\rm min}$ for an exposure of t seconds, given an 
IACT of effective area $A_{\rm eff}$, angular resolution $\sigma_{\rm t}$
and threshold energy $E_{\rm th}$ exceeds a significant number $M_s$ of 
standard 
deviations ($M_s\sigma$) the background noise $\sqrt{N_b}$, i.e.
$F_{\rm min}A_{\rm eff}t/\sqrt{N_b} \geq$ $M_s$ 
(see, e.g., \cite{Bergstrom98,TO}). The background counts ($N_b$) due to 
electronic and hadronic (cosmic protons and helium nucleids) cosmic ray
showers have been estimated using the 
following expressions \cite{Bergstrom98}:
$N_e = 3 \times 10^{-2} E_{\rm th}^{-2.3} \, t \, A_{\rm eff} \, \Delta\Omega,$
$N_h = 6.1 \times 10^{-3}  E_{\rm th}^{-1.7} \, t \, A_{\rm eff} \, \Delta\Omega$. As an additional background, one has to consider also the contamination
due to isolated muons which depending on the f.o.v. and altitude
location of the telescope may be even the dominant background at some
energy range (the ``muon wall''). Preliminary studies \cite{Manel} show
that the muon background could be as relevant as the hadronic background at
$E_{\rm th}\gtrsim100\,GeV$ while it can be effectively rejected at
lower $E_{\rm th}$. The diffuse galactic and 
extragalactic gamma radiation are negligible compare to this background. Gamma
point-like sources within the f.o.v can be resolved and 
taken out a posteriori.  The $E_{\rm th}$ 
of an IACT depends on the zenith angle of observation. The Galactic Center
is visible at different zenith angles by all present 
IACTs (e.g. CANGAROO-III, H.E.S.S.,MAGIC, VERITAS), but in the best case 
an $E_{\rm th}$ of about 100\,GeV can be achieved. Nevertheless, future
planed installations may reduce the $E_{\rm th}$ below 10\,GeV. The 
$A_{\rm eff}$ is also sensible to the zenith
angle of observation, here we choose a value of $1\times10^9$ cm$^{2}$.
This detectability condition will
allow us to compute the $5\sigma$ minimum detectable flux F$_{\rm min}$ 
in 250 hours of integration  with a typical IACT  
of $A_{\rm eff}=1\times10^9$ cm$^{2}$ and $E_{\rm th}=100\,$GeV (dashed 
line in Fig.~\ref{fig:fig2}). At a given distance from the Galactic Center 
only the flux values,
for a particular model of the Milky Way, greater 
than F$_{\rm min}$ will be detected. The detection will be much harder and
may result only in a central spot in the case of an IACT with
higher $E_{\rm th}$, as the YSUS parameter declines with $E_{\rm th}$
(see Fig.~\ref{fig:fig2}).

The Milky Way models presented here include adiabatic compression and
likely will result in a detection of the annihilation signal no matter
what was the initial (uncompressed) DM density profile. For current
experiments this detection will be successful only for the very
central regions, less than $\sim0.4^\circ$ in the case of the Model A
and close to $\sim1^\circ$ in the case of the Model B.  The compressed
Moore et al. DM profile will provide a more extended gamma flux
profile. The uncompressed NFW DM profile of the Model A will not be
detected even in the direction of the Galactic Center. On the
other hand, even the uncompressed Moore et al. profile of the Model B
will give a positive detection in the very inner regions of the Milky
Way. 

The effect of the adiabatic compression included in our Milky Way mass
models, which was previously ignored, is a crucial factor. It should
be emphasized that for the central $\sim 3$kpc of the Milky Way, where
the baryons dominate, it does not make sense to use the dark matter
profiles provided by cosmological N-body simulations: the DM
must fall into the deep potential well created by the collapsed
baryons. Thus, the models presented here are not extreme: they are the
starting point for realistic predictions of the annihilation
fluxes. One can envision few mechanisms to reduce the effect of the
compression. Transfer of the angular momentum to the dark matter as
suggested in \cite{Klypin02} is an option. Yet, recent simulations of
formation of bars indicate that it is difficult to arrange a
significant transfer of the angular momentum to the dark matter. The
DM density in the central few parsec can be reduced if
the central black hole formed by spiraling and merging of two black
holes \cite{Ulio}. It can also be changed (probably reduced) by
scattering of DM particles by stars in the central 2~pc
\cite{GnedinPrimack}. If this happens, the flux from the central 2~pc
can be significantly reduced. Yet, it will only decrease the amplitude
of the central spike. The signal from $0.4^\circ$ still could be
detected because it mostly comes from distances 30-50~pc, which are
much less affected by the uncertain physics around the black hole.

\begin{acknowledgments}
We acknowledge support of NASA and NSF grants to NMSU. We thank J. Primack, 
O. Gnedin, J. Betancort, A. Tasitsiomi and W. Wittek for 
discussions.
\end{acknowledgments}

\end{document}